\input psfig.sty
\documentstyle{cupconf}
\ifoldfss
\else
  \ifnfssone
    \newmathalphabet{\mathit}
      \addtoversion{normal}{\mathit}{cmr}{m}{it}
      \addtoversion{bold}{\mathit}{cmr}{bx}{it}
    \newmathalphabet{\mathcal}
      \addtoversion{normal}{\mathcal}{cmsy}{m}{n}
    \else
    \ifnfsstwo
    \fi
  \fi
\fi

\def\hexnumber#1{\ifcase#1 0\or1\or2\or3\or4\or5\or6\or7\or8\or9\or
 A\or B\or C\or D\or E\or F\fi }

\makeatletter
\ifx\CUP@mtlplain@loaded\undefined
\else
\fi
\makeatother
 \makeatletter
 \ifx\CUP@mtlplain@loaded\undefined
   \font\tenbmi=cmmib10 at 10pt
   \font\sevenbmi=cmmib10 at 7pt
   \font\fivebmi=cmmib10 at 5pt

   \newfam\bmifam
   \textfont\bmifam=\tenbmi
   \scriptfont\bmifam=\sevenbmi
   \scriptscriptfont\bmifam=\fivebmi
   
 \fi
 \makeatother
\ifnfsstwo

\fi
\ifnfssone

\fi
\ifoldfss

\fi

\mathchardef\varLambda="0103

\makeatletter
\ifx\CUP@mtlplain@loaded\undefined
\else
\fi
\makeatother
\makeatletter
\ifx\CUP@mtlplain@loaded\undefined
  \font\tenbms=cmbsy10
  \font\sevenbms=cmbsy10 at 7pt
  \font\fivebms=cmbsy10 at 5pt
  \newfam\bmsfam
  \textfont\bmsfam=\tenbms
  \scriptfont\bmsfam=\sevenbms
  \scriptscriptfont\bmsfam=\fivebms

  \edef\bsy@{\hexnumber\bmsfam}
  \mathchardef\bnabla="0\bsy@72
\fi
\makeatother
\def\etal{\mbox{\it et al.}}

\title[VLBI observation of five compact radio sources]{VLBI observations of five
compact radio sources}

\author[J. F. Zhou {\it et al.\/}]%
{J.\ls F.\ns Z\ls H\ls O\ls U$^1$$^2$,\ns D.\ls R.\ns J\ls I\ls
A\ls N\ls G$^1$$^2$,\ns X.\ls Y.\ns H\ls O\ls N\ls G$^1$$^2$\ns\\
\and \ns T.\ns  V\ls E\ls N\ls T\ls U\ls R\ls I$^3$}

\affiliation{$^1$Shanghai Astronomical Observatory, No. 80, Nandan Road,
Shanghai 200030, China\\[\affilskip]
$^2$National Astronomical Observatories, Chinese Astronomical Society, China\\[\affilskip]
$^3$Istituto di Radioastronomia del CNR, Via Gobetti 101, I-40129 Bologna, Italy
}

\begin{document}
\ifnfssone
\else
  \ifnfsstwo
  \else
    \ifoldfss
      \let\mathcal\cal
      \let\mathrm\rm
      \let\mathsf\sf
    \fi
  \fi
\fi

\maketitle

\begin{abstract}
Five compact radio sources, include 0420-014, 1334-127, 1504-166,
2243-123, and 2345-167, were observed at 5GHz by European VLBI
(Very Long Baseline Interferometry) Network (EVN) in June, 1996.
The primary purpose of this observation was to confirm their
superluminal proper motions. Here, the results of 1334-127,
1504-166, 2243-123 and 2345-167 are presented.
\end{abstract}

\firstsection % if your document starts with a section,
              % remove some space above using this command.
\section{Introduction}
The blazars exhibit the most rapid and the largest amplitude
variations of all AGN (Stein \etal\/ 1976; Angel \& Stockman
1980). Extreme variability suggests the continuum is emitted by a
relativistic jet close the line of sight and hence that the
observed radiation is strongly amplified by beaming effect
(Blandford \& Rees 1978).

In the radio band, blazars usually have very compact,
flat-spectrum cores, which are appropriate for VLBI observation.
The brightness temperature of a core can be estimated by VLBI
observation. Multi-frequency VLBI observation can obtain the
spectrum of a core as well as jet components. Also, if a jet
component shows superluminal proper motion, then it is possible to
estimate its Doppler boosting factor $\delta$ and  viewing angle
(Marscher 1987).

Shen et al. and Hong et al. carried three VLBI surveys of compact
radio sources on southern-hemisphere in November 1992 (Shen
\etal\/ 1997), May 1993 (Shen \etal\/ 1998) and October 1995 (Hong
et al. 1999). In these surveys, source 0420-014, 1334-127,
1504-166, 2243-123 and 2345-167 show some jet structures. The aim
of new EVN observations was to confirm the superluminal proper
motions in these sources. Here, we present some results of
1334-127, 1504-166, 2243-123 and 2345-167.  The results of
0420-014 were discussed in detail in Zhou \etal\/ (2000) because
of its notable interesting properties.

Throughout the paper, we assume $H_{\rm 0} = 100 h\; {\rm km\;
s^{-1}\; Mpc^{-1} }$ and $q_{\rm 0} = 0.5$.

\section{Observation and data reduction}

The VLBI observations were carried out from 1996 June 15 to 17
with EVN. Eight telescopes took part in the observations, i.e.
Shanghai, Crimea, Noto, Hartebeesthoek, Urumqi, Onsala, WSRT, and
Torun.  The recording modes was MkIII mode E (bandwidth 14MHz).
The data were correlated at MPIfR Mk III correlator in Bonn.

\begin{figure}
\centerline{ \psfig{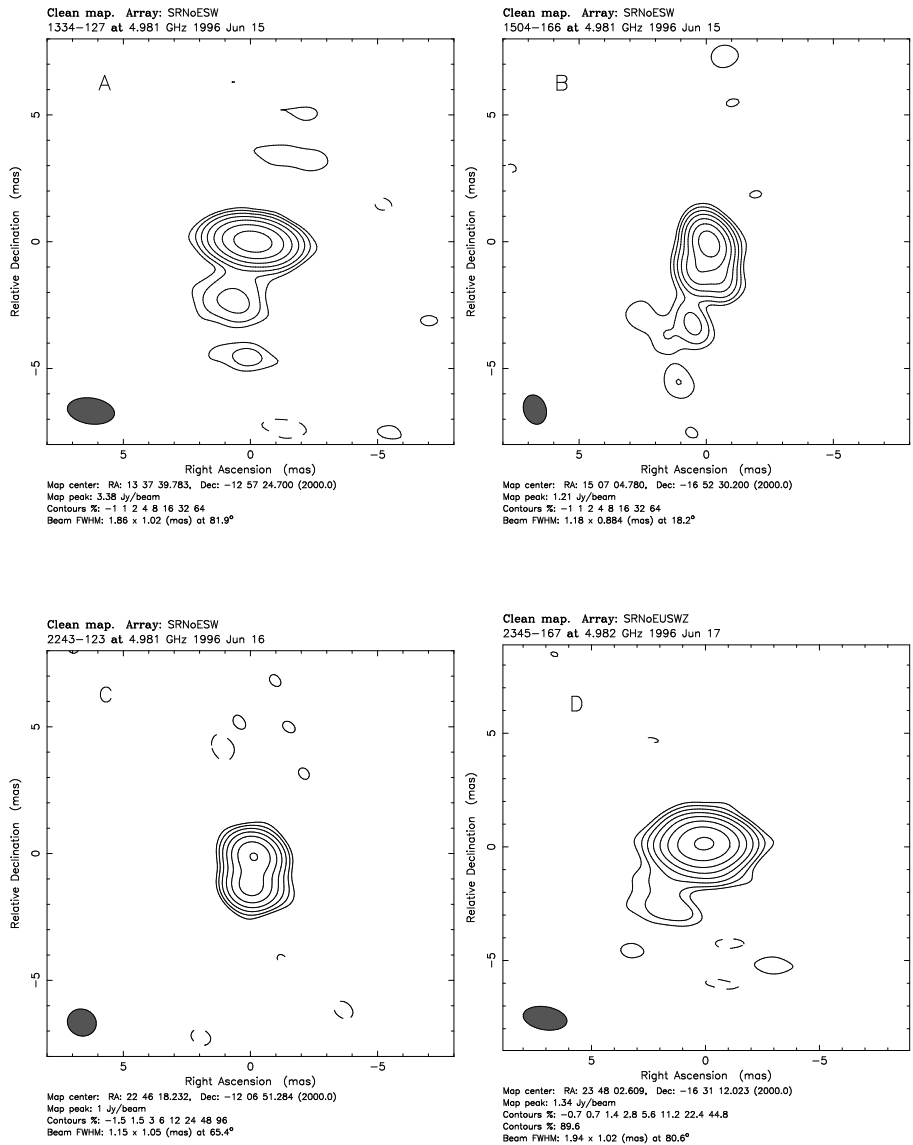} }
  \caption{CLEAN maps of 1334-127(A), 1504-166(B), 2243-123(C)
    and 2345-167(D). \label{fig1}}
\end{figure}

A-priori amplitude calibration and fringe-fitting were carried out
using the standard routines in the AIPS package. Imaging was done
using the DIFMAP difference mapping software (Shepherd, Pearson,
\& Taylor 1994). Four final CLEAN maps of 1334-127, 1504-166,
2243-123 and 2345-167 are displayed in Figure \ref{fig1}.  In
order to carry out the image analysis, we used the program
`modelfit' in DIFMAP to fit the final self-calibrated UV data with
circular gaussian components. The model fitting results are listed
in Table \ref{tbl-1}.

\section{Results and discussion of individual sources}

\begin{table}
  \begin{center}
  \begin{tabular}{ccccccc}
      Source & Component  & S   & Radius & P.A. & Major & Ratio \\
       Name  & Number     & (Jy) & (mas) & (deg) & (mas) &
       \\[3pt]

       1334-127 & 1 & 3.3 & 0 & 0 & $7.5e^{-5}$ & 1.0  \\
                & 2 & 0.4 & 2.5 & 159 & 0.65 & 1.0  \\[3pt]
       1504-166 & 1 & 1.2 & 0 & 0 & 0.39 & 1.0 \\
                & 2 & 0.7 & 1.3 & 164 & 0.94 & 1.0 \\[3pt]
       2243-123 & 1 & 0.7 & 0 & 0 & $2.5e^{-7}$ & 1.0  \\
                & 2 & 1.1 & 1.2 & -10 & 0.27 & 1.0 \\[3pt]
       2345-167 & 1 & 2.0 & 0 & 0 & 0.9 & 1.0 \\
                & 2 & 0.1 & 3.4 & 138 & 0.7 & 1.0 \\[3pt]
  \end{tabular}
  \caption{Model descriptions of 1334-127, 1504-166, 2243-123 and 2345-167.
            \label{tbl-1}}
  \end{center}
\end{table}

\subsection{1334-127}

PKS 1334-127 ( z = 0.539 ) was classified as a highly polarized
quasar by Impey \& Tapia (1988). It is also a ROSAT (Brinkmann,
Siebert, \& Boller 1994) and an EGRET (Hartman \etal\/ 1999)
source.

The large scale  VLA image at 1.4GHz shows a curved jet extending
6.5 arcseconds to the east of the core (Perley \etal\/ 1982). VLBI
observation carried out at 5GHz in 1986.9 (Wehrle \etal\/ 1992)
indicated that the source was barely resolved. A jet component was
detected at 1.7mas from the core in $P.A. = -150^\circ$ in
November 1992 (Shen \etal\/ 1997). In 1995, the jet component
moved to the position with radius of 2.11 mas and $P.A. \approx
170^\circ$ (Hong \etal\/ 1999). Shen \etal\/ (1997) reported a
proper motion of $0.28\; {\rm mas\, yr^{-1}}$ and Hong \etal\/
(1999) presented an apparent velocity of $\beta_{\rm app} =
2.4\pm0.5 h^{-1}$.

Our image, shown in Figure \ref{fig1}(A), indicates a jet
structure extending to south-east, with radius of 2.5 mas and
$P.A.$ of $159^\circ$ (Table \ref{tbl-1}). The jet is probably
bending clockwise. Thus, it is natural that substructures appear
in the east of large scale VLA map. With three epochs of VLBI
observation at 5GHz, we estimate a proper motion of $0.21\; {\rm
mas\, yr^{-1}}$ which corresponds to a apparent velocity of
$\beta_{\rm app} = 3.9 h^{-1}$.

\subsection{1504-166}
1504-166 (z=0.876) is classified as a highly polarized quasar
(Impey \& Tapia 1988), and is known to be a low frequency variable
source (McAdam 1982). It was unresolved at 1.4GHz by VLA (Perley
 \etal\/ 1982 ).

Two jet components were detected in the 5GHz VLBI observations in
1992 (Shen \etal\/ 1997). One is in the position with radius of
1.12 mas and P.A. of $-156^\circ$; the other is in the position
with radius of 0.8 mas and P.A. of $161^\circ$.  In the 5GHz VLBI
observation in 1995, Hong et al. (1999) detected a jet component
with radius of 1.35 mas and P.A. of $150^\circ$, which had a
proper motion of 0.08 ${\rm mas\, yr^{-1}}$.

Our VLBI image (Figure \ref{fig1}(B)) shows a slightly curved
core-jet structure to the south-east. The jet position angle is
consist with that derived at 1.7GHz by Romney (1984). The model
fitting results show a jet component in south-east direction
($P.A. = 164^\circ$). Assuming the jet component in our image is
the same one as the jet component in Hong et al. (1999) and the
component 2 in Shen et al. (1998), we can estimate a proper motion
of 0.055 ${\rm mas\, yr^{-1}}$, which corresponds to $\beta_{\rm
app} = 1.4 h^{-1}$.

\subsection{2243-123}

2243-123 (z=0.630) is a highly polarized quasar. Emission at X-ray
and $\gamma$-ray energies was detected (Maisack \etal\/ 1994;
Fichtel \etal\/ 1994). The VLA image shows an unresolved core and
an extended component at $4"$ from the core in $P.A. = 40^\circ$ (
Morganti \etal\/ 1993; Browne \& Perley 1986; Perley 1982). In the
15GHz (Kellermann \etal\/ 1998), 2.3GHz and 8.4GHz (Fey \& Charlot
2000 ) VLBA images, the jet clearly bends to the north-east
direction. Thus, the core is probably in the south.
 Hong \etal\/ (1999) suggests a proper motion of $0.40\; {\rm
mas\; yr^{-1}}$ assuming a jet component was ejected out in
1992.90.

Our new image (Figure \ref{fig1}(C)) shows a core-jet structure.
The two components are about 1.2 mas apart. Compare to the results
in 1995 (Hong \etal\/ 1999), a proper motion of $0.19\; {\rm mas\;
yr^{-1}}$ was estimated, which corresponds to an apparent velocity
$\beta_{\rm app} = 3.8 h^{-1}$. The jet component was probably
emerged in 1990 when a strong flare began.

\subsection{2345-167}

2345-167 (z=0.576) is an optically violent variable and a highly
polarized blazar. X-ray emission was detected by ROSAT (Brinkmann,
Siebert, \& Boller 1994). It has a complex radio spectrum with a
peak around 5GHz, and is a low frequency variable source (McAdam
1982). VLA observations show a jet of 4.0 arcseconds in $P.A. =
-130^\circ$. In milli-arcsecond scale, the jet moves out along the
direction of $\sim 110^\circ$. Therefore, Shen et al. (1997)
suggest that the jet may bend clockwise from south-east at
milli-arcsecond scale to south-west at arcseconds scale. Proper
motions of $0.26\; {\rm mas\; yr^{-1}}$ and $0.08\pm0.03\; {\rm
mas\; yr^{-1}}$ were reported by Shen \etal\/ (1997) and Hong
\etal\/ (1999) respectively.

Our image of 2345-167 shows a compact core and a weak jet
component in the south-east (Figure \ref{fig1}(D)). The jet
component is located about 3.4 mas from the core with $P.A. =
138^\circ$. Compare to the results of Shen et al. (1997) and Hong
et al. (1999), we obtain a proper motion of $0.085 {\rm mas\;
yr^{-1}}$. The corresponding apparent velocity is $\beta_{\rm
app}\approx 1.7\; h^{-1}$, which is comparable to Hong's (1999)
results.

%\section{Conclusions}
%The four sources discussed above all show superluminal proper
%motions. However, high resolution and high dynamic VLBI
%observations are needed to further study their sub-mas structures
%and the connection between mas and arcsecond structures.

\begin{acknowledgments}
This research has been supported by the Major Basic Research
Development Program and NNSFC grant 19773019.
\end{acknowledgments}

\end{document}